\documentclass{elsart}

\usepackage{epsfig}

\begin{document}

\begin{frontmatter}

\title{The validity of Cassie's law: A simple exercise using 
a simplified model.}

\author{Masao Iwamatsu\thanksref{label1}}
\thanks[label1]{ E-mail: iwamatsu@ph.ns.musashi-tech.ac.jp, Tel: +81-3-3703-3111 ext.2382, 
Fax: +81-3-5707-2222}
\address{Department of Physics, General Education Center,
Musashi Institute of Technology,
Setagaya-ku, Tokyo 158-8557, Japan }

\begin{abstract}
The contact angle of a macroscopic droplet on a heterogeneous but flat substrate is studied using the interface displacement model which can lead to the augmented Young-Laplace equation. Droplets under the condition of constant volume as well as constant vapor pressure are considered.  By assuming a cylindrical liquid-vapor surface (meniscus) and minimizing the total free energy of the interface displacement model, we derive an equation which is similar but different from the well known Cassie's law.  Our modified Cassie's law is essentially the same as the formula obtained previously by Marmur [J.Colloid Interface Sci. {\bf 168}, 40 (1994)].  A few consequences from this modified Cassie's law will be briefly described in the following sections of this paper.  Several sets of recent experimental results seem to support the validity of our modified Cassie's law.

\end{abstract}

\begin{keyword}
wetting \sep Cassie's law \sep contact angle \sep heterogeneous surface

\PACS 68.08.Bc \sep 05.20.Jj 
\end{keyword}
\end{frontmatter}

\section{Introduction}

Cassie's law~\cite{Cassie1,Cassie2} is for the contact angle~\cite{Young} of a macroscopic droplet on a chemically heterogeneous surface and has recently attracted attention for its potential application to the design issue of the microfluidic devices in chemical and biomedical engineering.  For example, this law is successfully used to explain the superhydrophobicity and self-cleaning mechanism of various natural and artificial surfaces~\cite{Quere}.

Although the validity of Cassie's law is conceptually verified using various theoretical models~\cite{Swain,Henderson,Rascon} and computer simulations using the molecular dynamics~\cite{Adao} and the Monte Carlo method~\cite{Urban}, some model calculations ~\cite{Israelachivili,Marmur} as well as numerical simulations using lattice Boltzmanm~\cite{Zhang} and spin system~\cite{Pesheva} raised questions about the validity of Cassie's law.  This doubt is further reinforced by various recent experimental results~\cite{Woodward,Cubaud} which became possible through the recent progress of micro and nano-scale fabrication of artificial surfaces.

In this paper, we will consider a simplified model of translationally symmetric cylindrical droplet on a heterogeneous but smooth surface, and derive the expressions for the apparent contact angle of the droplet.  To this end, we will take the continuum model and use the framework of the so-called interface displacement model~\cite{Swain,Brochard}.  By solving the augmented Young-Laplace equation derived from the interface displacement model, and minimizing the model free energy for a cylindrical droplet, we will derive a formula for the contact angle of the droplet on a heterogeneous surface.  The formula obtained, which we called the modified Cassie's law, is similar but clearly different from the original Cassie's law.  Our modified Cassie's law states that the average of the local contact angle along the edge of droplet will be the observable apparent contact angle rather than the average over the contact area of the droplet. Several recent experimental works~\cite{Woodward,Cubaud} support our modified Cassie's law.

The format of this paper is as follows: section2 will be a review of Cassie's law and the interface displacement model~\cite{Swain,Brochard} which leads to the augmented  Young-Laplace equation.  In section 3, a simplified model will be used to derive the expression for the apparent contact angle.  Section 4 of this paper is devoted to the concluding remarks.

\section{Cassie's law and the interface displacement model for a heterogeneous surface}

Cassie~\cite{Cassie1,Cassie2} has shown that by averaging the surface energy, the contact angle $\theta_{C}$ of a heterogeneous surface, which consists of two types of surfaces with contact angle $\phi_{1}$ with area fraction $c_{1}$ and $\phi_{2}$ with areal fraction $c_{2}$ is given by
\begin{equation}
\cos\theta_{C} = c_{1}\cos\phi_{1}+c_{2}\cos\phi_{2}
\label{eq:2-1}
\end{equation}
According to this Cassie's law, the apparent contact angle $\theta_{C}$ of a droplet on a heterogeneous surface with spatially inhomogeneous local contact angle $\phi(x)$ is given by the areal average of the local contact angle
\begin{equation}
\cos\theta_{C} = <\cos\phi(x)>
\label{eq:2-2}
\end{equation}
where $<>$ denote the areal average given by the areal integral of the cosine $\cos\phi(x)$ of the local contact angle $\phi(x)$.

Since, eq.(\ref{eq:2-1}) and (\ref{eq:2-2}) is derived from the thermodynamic
definition of contact angle~\cite{Cassie2}, the angle $\theta_{C}$ is {\it not} an
observable apparent contact angle but is really a conceptual effective contact angle. Therefore,
eqs.(\ref{eq:2-1}) or (\ref{eq:2-2}) is not an equation to calculate the
apparent contact angle but is rather the {\it definition} of the effective
contact angle $\theta_{C}$.  Therefore, the contact angle $\theta_{C}$ calculated
from Cassie's law cannot be
used to interpret the observed real contact angle on a heterogeneous surface.
It can be compared only to the ensemble average of the 
cosines of many contact angles at
different postions on a heterogeneous surface~\cite{Swain}.  The Cassie's law is quite
general in a sense that it can describe the surface free enegy of adsorbed
liquid films on a heterogeneous surface well~\cite{Swain,Henderson,Rascon}, but it should not be connected to the contact angle as Cassie himself did~\cite{Cassie2}.
This fact
seems to cause many confusions and misunderstandings of the validity of the
Cassie's law~\cite{Zhang,Pesheva}, which we will reconsider in this paper.

In order to verify this Cassie's law as a formula to predict the observed apparent contact angle on a heterogeneous surface, we will use the so-called interface displacement model~\cite{Swain,Brochard} and the corresponding augmented Young-Laplace equation.  The free energy functional $\mathcal{F}$ of wetting layer with thickness $z(x)$ on a smooth surface along $x$ axis in an environment of undersaturated
vapor pressure $p_{0}$ is given by
\begin{equation}
\mathcal{F} = \int F\left(z,z'\right) dx
\label{eq:2-3}
\end{equation}
where $z'=dz/dx$.  The local free energy functional $F$ is given by
\begin{equation}
F = \sigma_{lf}\sqrt{1+\left(\frac{dz}{dx}\right)^{2}}
+\left(\sigma_{sl}(x)-\sigma_{sf}(x)\right)+P(z,x)-p_{0}z 
\label{eq:2-4}
\end{equation}
where, the pressure $p_{0}$ is measured relative to the saturated vapor pressure, so that the saturated vapor pressure corresponds to $p_{0}=0$.  The constant pressure ensemble of statistical mechanics was used as a natural choice which corresponds to the real experimental situation.

$\sigma_{lf}$ is the liquid-fluid(vapor) and $\sigma_{sl}(x)$ is the solid(substrate)-liquid surface tensions.  Similarly $\sigma_{sf}(x)$ denotes the solid(substrate)-fluid(vapor) surface tension.  The heterogeneity comes in only through the positional dependence of these two surface tensions $\sigma_{sl}(x)$ and $\sigma_{sf}(x)$. The finite thickness $z(x)$ of the liquid-solid interfacial energy is further corrected by the thin-film free 
energy $P(z,x)$, which is related to the so-called disjoining pressure $\Pi(z,x)$~\cite{Davis} by
\begin{equation}
\Pi(z,x) = -\frac{\partial P}{\partial z}
\label{eq:2-5}
\end{equation}
This thin-film energy plays a crucial role in surface phase transition~\cite{Dietrich}, and the microscopic real contact angle of the droplet which will not be covered in this paper since we are mainly concerned with a macro-scale or micro-scale droplet~\cite{Dietrich2}.  
From now on, we will neglect the thin film force $P(x)$ and set $P(x)=0$.
The meniscus of translationally symmetric cylindrical droplet is
determined from the Euler-Lagrange equation:
\begin{equation}
\frac{d}{dx}\left(\frac{\partial F}{\partial z^{'}}\right)-\frac{\partial F}{\partial z}=0
\label{eq:2-6}
\end{equation}
which leads to
\begin{equation}
\frac{1}{\left[1+(dz/dx)^{2}\right]^{3/2}}\frac{d^{2}z}{dx^{2}}-p_{0}=0
\label{eq:2-7}
\end{equation}
where we have neglected the thin-film energy $P(x)$.  The differential equation (\ref{eq:2-7}) has a solution, which is a semicircle~\cite{Davis} with the radius $R$ given by
\begin{equation}
R=\frac{\sigma_{lf}}{p_{0}}
\label{eq:2-8}
\end{equation}
Therefore, the radius of curvature $R$ of the meniscus of the droplet is determined from the pressure $p_{0}$ which is also called the capillary pressure.  One should note that the heterogeneity and the site dependence of the surface tensions $\sigma_{sl}(x)$ and $\sigma_{sf}(x)$ does not play any role in the global shape of the meniscus of macro- and micro-droplets.

\section{Cylindrical droplet on a heterogeneous surface}

Experimentally, one might consider the two types of droplets on a heterogeneous surface: the one in which the volume of the droplet can be fixed externally, and another in which the vapor pressure of the environment can be controlled.  The former belongs to the ensemble of the constant volume and the latter to the constant pressure.  Although the thermodynamic functions are related each other through the Legendre transform, we will consider the two casese separately because they represent different experimental situations.

\subsection{Drop of constant volume}

We considered the cylindrical droplet of base length $2r$ whose center is fixed on a heterogeneous surface.  The vapor pressure is saturated pressure, so that $p_{0}=0$.  Choosing the coordinate system shown in Fig. 1, we have the total free energy:
\begin{equation}
\mathcal{F} = \sigma_{lf}\frac{2r\theta}{\sin\theta}
+\int_{-r}^{r}\left(\sigma_{sl}(x)-\sigma_{sf}(x)\right)dx
\label{eq:3-1}
\end{equation}
which should be minimized with respect to the apparent contact angle $\theta$ subject to the condition of the constant volume $S$ (which is actually the cross section)
\begin{equation}
S = \int_{-r}^{r}z(x)dx=r^{2}\frac{\theta-\sin\theta\cos\theta}{\sin^{2}\theta}
\label{eq:3-2}
\end{equation}
which gives the relationship between the base length $r$ and the contact angle $\theta$:
\begin{equation}
\frac{dr}{d\theta}=r\frac{\theta\cot\theta-1}{\theta-\sin\theta\cos\theta}
\label{eq:3-3}
\end{equation}

\begin{figure}[htbp]
\begin{center}
\includegraphics[width=0.8\linewidth]{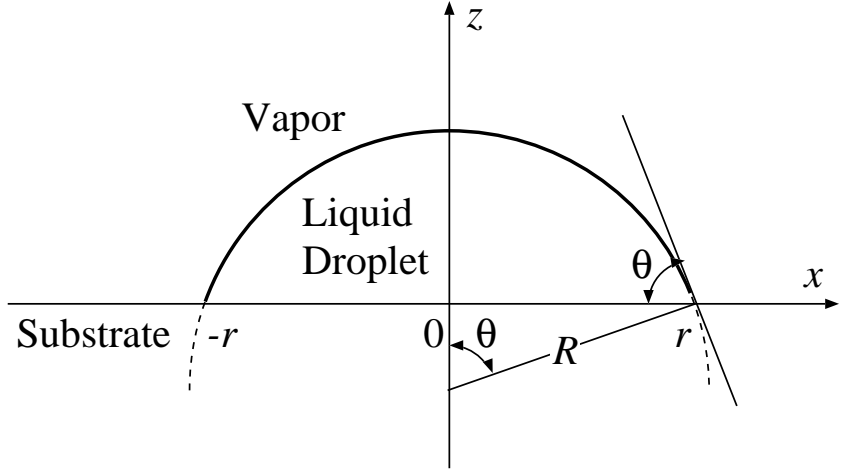}
\caption{A droplet on a heterogeneous surface.  The center of the droplet is fixed at the origin of the $x$ coordinate.  The base length is $2r$ which is related to the radius of curvature of the meniscus $R$ through $r=R\sin\theta$ where $\theta$ is the apparent contact angle. }
\label{Fig:1}
\end{center}
\end{figure}

Now, writing the surface free energy using the local contact angle $\phi(x)$ defined by the Young's equation~\cite{Young}:
\begin{equation}
\sigma_{sl}(x)-\sigma_{sf}(x)=-\sigma_{lf}\cos\phi(x)
\label{eq:3-4}
\end{equation}
we find
\begin{equation}
\mathcal{F} = \sigma_{lf}\frac{2r\theta}{\sin\theta}
-\sigma_{lf}\int_{-r}^{r}\cos\left[\phi(x)\right]dx
\label{eq:3-5}
\end{equation}
Differentiation of eq.(\ref{eq:3-5}) with respect to $\theta$, and taking into account the fact that $r$ is also the function of $\theta$ through (\ref{eq:3-3}), we arrive at the condition $d\mathcal{F} /d\theta=0$ to minimize the total free energy (\ref{eq:3-1}):
\begin{equation}
\cos\theta = \frac{1}{2}\left(\cos\phi(r)+\cos\phi(-r)\right)
\label{eq:3-6}
\end{equation}
Therefore, the apparent (actual) contact angle $\theta$ is the average of the contact angle at the left and the right edge of the droplet:
\begin{equation}
\cos\theta = \overline{\cos\phi}
\label{eq:3-7}
\end{equation}
where
\begin{equation}
\overline{\cos\phi}=\frac{1}{2}\left(\cos\phi(r)+\cos\phi(-r)\right)
\label{eq:3-8}
\end{equation}
which is similar, but different from Cassie's law.  This  Cassie-like law, which we will call the modified Cassie's law, states that the cosine of the apparent contact angle $\theta$ is give by the average of the cosine of the local contact angle $\phi(-r)$ at the left edge and $\phi(r)$ at the right edge.  Therefore the apparent contact angle cannot be obtained from the areal average of the cosine of the local contact angle, rather it will be obtained from the average along the edge of the droplet.  Essentially the same result was obtained by Marmur~\cite{Marmur} using a more restricted model.

When the radius of curvature $R$ is infinitely large, we can neglect the first term of (\ref{eq:3-1}), and the total free energy is approximated by
\begin{equation}
\mathcal{F} \simeq -\sigma_{lf}\int_{-r}^{r}\cos\phi(x) = -2r\sigma_{lf}<\cos\phi(x)>
\label{eq:3-9}
\end{equation}
where $<\cos\phi(x)>$ is the areal average, which can be expressed using Cassie's law as
\begin{equation}
\mathcal{F} \simeq -2r\sigma_{lf}\cos\theta_{C}
\label{eq:3-10}
\end{equation}
where we have used the definition of Cassie's angle $\theta_{C}$ given by eq. (\ref{eq:2-2}).  Then, the free energy is minimized when the cosine of the contact angle calculated from Cassie's law is maximized.

If the droplet can move freely along the surface, one should minimize the free energy of droplet and this is given by
\begin{equation}
\mathcal{F} = \sigma_{lf}\frac{2r\theta}{\sin\theta}
+\int_{X}^{X+2r}\left(\sigma_{sl}(x)-\sigma_{sf}(x)\right)dx
\label{eq:3-11}
\end{equation}
instead of (\ref{eq:3-1}).  Now, not only the apparent contact angle $\theta$ but the position $X$ of the droplet (see Fig. 2) is variable.  

\begin{figure}[htbp]
\begin{center}
\includegraphics[width=0.8\linewidth]{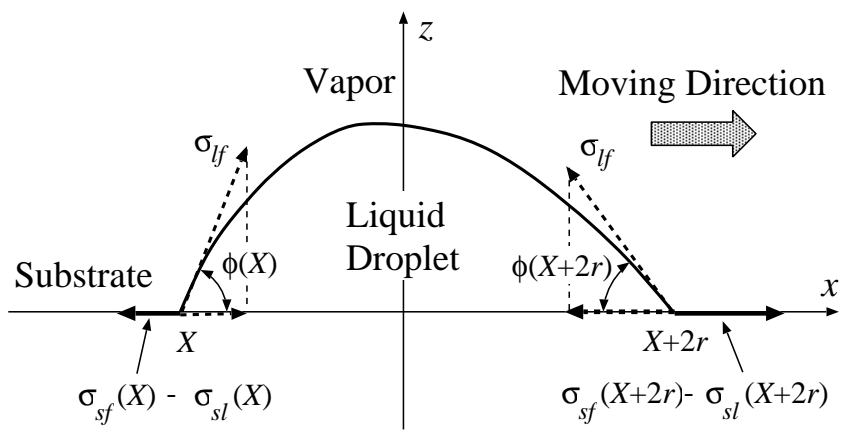}
\caption{ A freely moving droplet on a heterogeneous surface.  The left edge of the droplet is at $X$ and the right edge at $X+2r$. The droplet can move freely along the $x$ coordinate due to the force imbalance at the left edge
from the substrate $\sigma_{sf}(X)-\sigma_{sl}(X)$ and at the right edge
from the substrate $\sigma_{sf}(X+2r)-\sigma_{sl}(X+2r)$.
}
\label{Fig:2}
\end{center}
\end{figure}

By minimizing (\ref{eq:3-11}) with respect to $\theta$ and $r$ through (\ref{eq:3-3}), one obtains
\begin{equation}
\cos\theta = \cos\phi(X+r)
\label{eq:3-12}
\end{equation}
This equation should be augmented by the condition that the position of the droplet becomes the minimum of the free energy.  This condition is derived by minimizing the above free energy (\ref{eq:3-11}) with respect to the position $X$, which leads to
\begin{equation}
\cos\phi(X+2r)=\cos\phi(X)
\label{eq:3-13}
\end{equation}
Combining eqs.(\ref{eq:3-12}) and (\ref{eq:3-13}), we arrive at the modified Cassie's law (\ref{eq:3-8})
\begin{equation}
\cos\theta = \frac{1}{2}\left(\cos\phi(X)+\cos\phi(X+2r)\right)
\label{eq:3-14}
\end{equation}
again.  However the contact angle $\phi(X)$ at the left and $\phi(X+2r)$ at the right edge is always equal in this case.

When the droplet can move freely, it will move to a position where the local contact angle $\phi(X)$ of the left edge and $\phi(X+2r)$ of the right edge are equal.  This condition has a simple mechanical meaning since eq. (\ref{eq:3-13}) can be written as
\begin{equation}
\sigma_{sf}(X+2r)-\sigma_{sl}(X+2r)
= \sigma_{sf}(X)-\sigma_{sl}(X)
\label{eq:3-15}
\end{equation}
from the Young's equation (\ref{eq:3-4}).  Since eq. (\ref{eq:3-15}) represents the balance of the force exerted from the substrate at the left and the right edge of the droplet as shown in Fig. 2, the minimum total free energy implies the force balance.  This is the natural condition of the static (non moving) droplet.  Therefore a thermodynamic equilibrium is realized when the mechanical balance is satisfied.

When the force balance is violated, a net force from the substrate remains and acts on the droplet. Then, the droplet will be dragged along the surface by the substrate until it moves to the position where the two forces at both edges cancel each other out.  This force imbalance could be macroscopic.
Suppose, the surface tension of the liquid-vapor interface $\sigma_{lf}$ is 10 mN/m. Then the force imbalance could be the same order and the net force $F$ act on the edge of length 1 mm is 10$^{-5}$N. Suppose the droplet is a cube with a side length of 1 mm and whose density is 1 g/cm$^{3}$.  Then the mass of the cube is 10$^{-6}$ kg, and the acceleration $a$ of the drop can be $a=10^{-5}$N/10$^{-6}$kg=10m/s$^{2}$, which is the same order of magnitude of gravitational acceleration $g$.  Therefore, if we can realize an alternating surface tension $\sigma_{sl}$ and $\sigma_{sf}$ in some way, we can easily move and guide the micro-droplet along the surface.

\subsection{Drop in the undersaturated vapor}

Next, we consider the cylindrical droplet whose center is fixed on a heterogeneous surface again, but the vapor pressure is under-saturated pressure $p_{0}\neq 0$.  Then, the volume of the droplet cannot be fixed but is determined from the vapor pressure $p_{0}$.  In this case, the radius of the curvature $R$ of meniscus is given by (\ref{eq:2-8}) and is fixed from the vapor pressure $p_{0}$.

Assuming the configuration shown in Fig.1, we have the total free energy:
\begin{equation}
\mathcal{F} = 2\sigma_{lf} R\theta
+\int_{-R\sin\theta}^{R\sin\theta}\left(\sigma_{sl}(x)-\sigma_{sf}(x)\right)dx
-p_{0}R^{2}(\theta-\sin\theta\cos\theta)
\label{eq:3-16}
\end{equation}
from (\ref{eq:2-3}) and (\ref{eq:2-4}), where we have used the fact that the total volume $S$ is given by
\begin{equation}
S=\int_{-R\sin\theta}^{R\sin\theta}z(x)dx = R^{2}(\theta-\sin\theta\cos\theta)
\label{eq:3-17}
\end{equation}

Equation (\ref{eq:3-16}) should be minimized with respect the actual contact angle $\theta$ subject to the condition of the constant curvature $R$ instead of the constant volume $S$ of the previous case.

From the expressions for the curvature $R$ in (\ref{eq:2-8}), and the definition of the local contact angle $\phi(x)$ in (\ref{eq:3-4}) we have the free energy
\begin{equation}
\mathcal{F} = \sigma_{lf}R(\theta+\sin\theta\cos\theta)
-\sigma_{lf}\int_{-R\sin\theta}^{R\sin\theta}\cos\left[\phi(x)\right]dx
\label{eq:3-18}
\end{equation}
By minimizing this free energy $\mathcal{F}$ with respect to the contact angle $\theta$ with $R$ as a constant, we have
\begin{equation}
\cos\theta=\frac{1}{2}\left(\cos\phi(R\sin\theta)+\cos\phi(-R\sin\theta)\right)
\label{eq:3-19}
\end{equation}
Therefore, the cosine of the apparent (actual) contact angle $\theta$ is the average of the cosine of the local contact angle $\phi(-R\sin\theta)=\phi(-r)$ at the left and $\phi(R\sin\theta)=\phi(r)$ at the right edge of the droplet:
\begin{equation}
\cos\theta = \overline{\cos\phi}
\label{eq:3-20}
\end{equation}
where $\overline{\cos\phi}$ is given by (\ref{eq:3-8}).  So, we can recover the modified Cassie's law (\ref{eq:3-7}) again for the constant volume case as well as the constant pressure case.

We have concentrated on the problem of a droplet whose center is fixed.  The droplet of the free moving droplet can be treated similarly, and we can obtain the same results as the ones derived for the constant volume case.  The only difference here is that the volume cannot be changed directly by injecting the liquid externally, but it will be controlled by the vapor pressure $p_{0}$.

So far, we have used the interface displacement model, where a sharp interface is assumed~\cite{Dietrich}.  In fact, the density of the liquid droplet changes continuously from the liquid to the vapor phase in the interfacial region.  The width $w$ of the liquid-vapor interface which is defined as the thickness of the interfacial region is in the order of nanometer~\cite{Dietrich} in normal conditions.  Therefore, so long as the heterogeneity of the surface is in the order of micrometer, the above discussion will be valid and our modified  Cassie's law (\ref{eq:3-6}) and (\ref{eq:3-14}) should be used.

\begin{figure}[htbp]
\begin{center}
\includegraphics[width=0.65\linewidth]{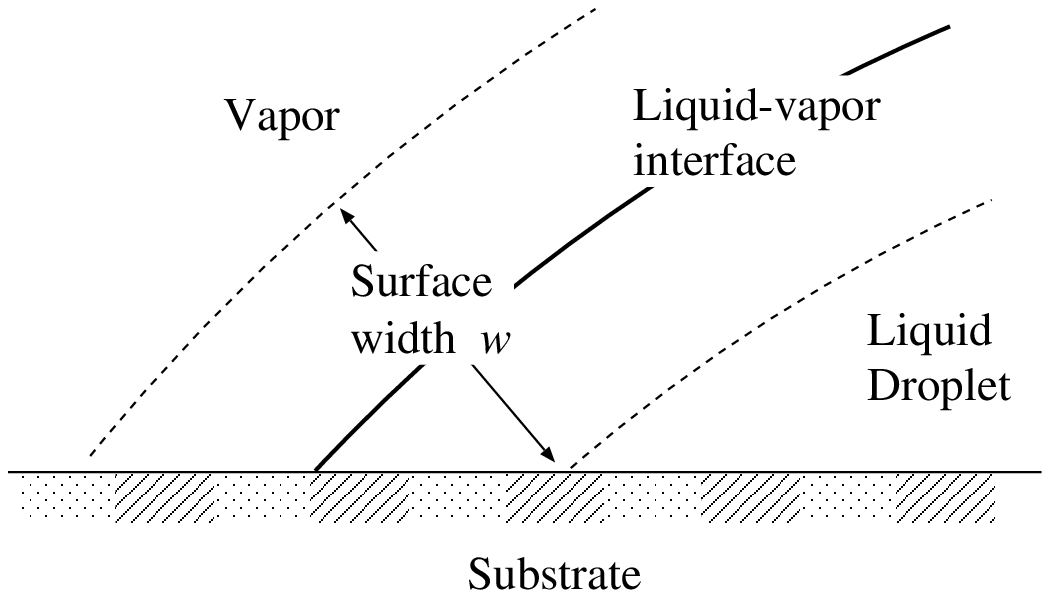}
\caption{
A droplet on a nanoscopic heterogeneous surface.  The striped surface is considered.  The width of the stripe is narrower than the liquid-vapor interface width $w$.  The region sandwiched by the two dashed lines is the interfacial region where the density changes continuously.  The solid line indicates the rough location of the liquid-vapor interface used in the interface displacement model.
}
\label{Fig:3}
\end{center}
\end{figure}

However, once the surface heterogeneity becomes nanoscopic, we cannot ignore the continuous density variation of the liquid droplet within the interfacial region.  Figure \ref{Fig:3} shows such a situation where the heterogeneous striped surface is considered.  Since the width of the stripe is narrower than the width $w$ of the liquid-vapor interface, we cannot define the position of the interface $z(x)$ precisely.  Then, it is intuitively natural to define the cosine of the apparent contact angle $\theta$ as the average of the cosine of the local contact angel $\phi(x)$ over the interfacial width $w$.  Then the apparent contact angle $\theta$ will be approximated by the original Cassie's law (\ref{eq:2-2}). The same conclusion was reached by Zhang and Kwok~\cite{Zhang}, who used lattice-Boltzmann model to calculate the contact angle of a cylindrical droplet on a striped surface.  Only in this case, the original Cassie's law can be interpreted as a formula to describe the observed apparent contact angle. Otherwise, it
should not be used to predict and calculate the observed contact angle on a heterogeneous
surface.  The Cassie's law is valid as a formula to calculate the free energy of
liquid on a heterogeneous surface, but its interpretation as an observable contact
angle on a heterogeneous surface is only valid for molecular-size heterogeneities.

\section{Concluding Remarks}
\label{sec:sec4}

In this paper, we used a simplified model to calculate the apparent contact angle of a cylindrical droplet under the condition of the constant volume and constant vapor pressure.

In  both cases, we found the same expression for the apparent contact angle akin to the Cassie's law.  The modified Cassie's law (\ref{eq:3-8}) and (\ref{eq:3-14}) we found requires the average of the local contact angle along the edge of the droplet rather than the contact area of the droplet.  We also studied a droplet which can move freely along the surface.  By minimizing the total free energy, we found not only the same modified Cassie's law but a condition that the local contact angle of the left and the right edge of the droplet must be equal.  The latter condition expresses the force balance acting on the droplet from the substrate.  Thus the local contact angle of the left and the right edge should always be equal and is also equal to the apparent contact angle observed. This force balance condition suggests the possibility of moving and driving micro-droplets along the surface.  The original Cassie's law (\ref{eq:2-2}) as a formula to calculate
the free energy is valid, but itd interpretsyion as a formula to calculate the observed
contact angle is valid only when the lenght scale of the heterogeneity of surface 
is shorter than the interfacial width of the liquid-vapor interface.

Although the statistical thermal average~\cite{Swain} and thermodynamic definition~\cite{Henderson} of the contact angle naturally obeys the original Cassie's law (\ref{eq:2-2}), the apparent contact angle, which is observed directly from the experiment on real macro- and micro-droplets, would satisfy our modified Cassie's law.  Several recent experimental results~\cite{Woodward,Cubaud} for axisymmetric hemispherical droplets are consistent with our modified Cassie's law. 
They~\cite{Woodward,Cubaud} found that the apparent contact angle satisfies the modified Cassie's law where the cosine of the apparent contact angle could be calculated from the average of the cosine of the local contact angle along the contour line rather than the average about the contact area.

In our report, we were concentrated on the macro- and micro-scopic droplet, and the macroscopic apparent contact angle is considered, which should be deduced from the extrapolation of the meniscus and could be observed using a microscope.  In order to study the nano-scopic contact angle of a nano-scopic droplet~\cite{Dietrich2}, which could be observed, for example, using the atomic force microscope (AFM)~\cite{Butt},
we have to include the thin film potential $P(x)$.  There are also the problem of line tension~\cite{Drelich}, which further modify the modified Cassie's law for the nanoscopic contact angle.  These issues are left for the future as well as other experimental investigations.

\section*{Acknowledgements}
The author is grateful to one of the reviewers for several important comments
which were very helpful to shape the final version of the manuscript.

\end{document}